\documentclass[aps,twocolumn,prd,showpacs,nofootinbib]{revtex4}
\usepackage{amsmath}
\usepackage{graphicx}
\usepackage{dcolumn}
\usepackage{bm}
\usepackage{amssymb}
\usepackage{latexsym}

\def\be{\begin{equation}}
\def\ee{\end{equation}}
\def\ba{\begin{eqnarray}}
\def\ea{\end{eqnarray}}

\begin{document}

\title{Adiabatic Spectra During Slowly Evolving}

\author{Yun-Song Piao}

\affiliation{College of Physical Sciences, Graduate University of
Chinese Academy of Sciences, Beijing 100049, China}

\begin{abstract}

In general, for single field, the scale invariant spectrum of
curvature perturbation can be given by either its constant mode or
its increasing mode. We show that during slowly expanding or
contracting, the spectrum of curvature perturbation given by its
increasing mode can be scale invariant. The perturbation mode can
be naturally extended out of horizon, and the amplitude of
perturbation is consistent with the observations. We briefly
discuss the implement of this scenario.

\end{abstract}

\maketitle

The nearly scale invariance of curvature perturbation is required
by the observations. How obtaining it is still a significant
issue, especially for single field. In general, the equation of
motion of curvature perturbation $\zeta$ in $k$ space is given by
\cite{Muk},\cite{KS}\footnote{ Eq.(1) is valid only for the case
with constant sound speed $c_s^2$, which we require here. When
$c_s^2$ is changed, the results obtained can be different from
that showed here. However, the essential how finding the solutions
generating the scale invariant spectrum is same as the discussion
given here. The reason that we only care the case with constant
$c_s^2$ is in principle the changes of all $a$, $|\epsilon|$ and
$c_s^2$ affect the spectrum, however, when both of them or all,
especially for $c_s$, are changed, the case is slightly
complicated for studying, thus the most interesting case might be
that one among them is changed, e.g. inflation.} \be
u_k^{\prime\prime} +\left(k^2-{z^{\prime\prime}\over z}\right) u_k
= 0 ,\label{uk}\ee where $'$ is derivative for $\eta=\int dt/a$,
$u_k \equiv z\zeta_k$ and $z\simeq a\sqrt{|\epsilon|}$, in which
$\epsilon=-{{\dot H}\diagup H^2}$ and $M_P^2=1$ is set. When
$k^2\ll z^{\prime\prime}/z$, the solution of Eq.(\ref{uk}) is
$\zeta_k\simeq C+D\int {d\eta\over z^2}$, e.g.\cite{Mukhanov}, in
which $C$ mode is the constant mode, and $D$ mode evolves with the
time.

In general, the scale invariance of spectrum requires
${z^{\prime\prime}\over z}\sim {2\over (\eta_*-\eta)^2}$. In
principle, both $a$ and $|\epsilon|$ can be changed, and together
contribute the change of $z$. When both are changed, the case is
complicated for studying. The simplest case is that one of both is
changed while another is hardly changed. When $|\epsilon |$ is
nearly constant, \be a\sim {1\over \eta_*-\eta}\,\,\, \text{or}
\,\,\, (\eta_*-\eta)^2 \label{a1}\ee have to be satisfied, where
initially $\eta\ll -1$ and $\eta_*$ is around the ending time of
the corresponding evolution. The evolution with $a\sim {1/
(\eta_*-\eta)}$ is that of the inflation
\cite{G},\cite{LAS},\cite{S1},\cite{MC}, in which $\epsilon \simeq
0$. While another is that of the contraction with $\epsilon\simeq
1.5$ \cite{Wands99},\cite{FB},\cite{SS}, both are dual
\cite{Wands99}. The increasing or decaying of $D$ mode is
determined by \be \int {d\eta\over z^2}\sim \int {d\eta\over
a^2{|\epsilon |}}\sim \left({\eta_*-\eta}\right)^{1-p},
\label{z}\ee where ${a^2{|\epsilon |}}\sim {(\eta_*-\eta)^p}$ is
applied. Thus it is decaying for $p<1$ and is increasing for
$p>1$. In general, for inflation, $D$ mode is decaying, the
spectrum is determined by the constant mode, since $p<1$, while
for the contraction with $\epsilon\simeq 1.5$,  the spectrum is
determined by the increasing $D$ mode, since $p>1$.

Thus the scale invariance of $\zeta$ can be given by either its
constant mode or its increasing mode, and the different modes
implies different scenarios of early universe. In principle, when
$|\epsilon |$ is nearly constant, the increasing mode of metric
perturbation $\Phi$, which is scale invariant for $\epsilon\gg 1$
or $\epsilon \ll -1$, might dominate the curvature perturbation.
The evolution with $\epsilon\gg 1$ corresponds to the slowly
contracting, which is that of ekpyrotic universe \cite{KOS}, while
$\epsilon \ll -1$ is the slowly expanding \cite{PZhou}, which has
been applied for island cosmology \cite{island}. The constant mode
of $\Phi$ is same with the constant mode of $\zeta$. The duality
of scale invariant spectrum of $\Phi$ has been discussed in
\cite{LST},\cite{Piao0404},\cite{Lid}. However, whether the
spectrum of $\zeta$ for $|\epsilon|\gg 1$ is scale invariant
depends that the increasing mode of $\Phi$ can be inherited by the
constant mode of $\zeta$ after exiting, which depends of the
physics around the exiting and is uncertain \cite{DH}.

When $|\epsilon |$ is rapidly changed while $a$ is nearly
constant, the scale invariance of $\zeta$ requires \be {|\epsilon
|}\sim {1\over (\eta_*-\eta)^2}\,\,\, \text{or} \,\,\,
(\eta_*-\eta)^4. \label{A}\ee It is found with Eq.(\ref{z}) that
for ${|\epsilon |}\sim (\eta_*-\eta)^4$, the scale invariance of
spectrum is determined by the increasing $D$ mode, while another
is determined by the constant mode. Eq.(\ref{A}) is only the
exchange of $\sqrt{|\epsilon |}$ with $a$ in Eq.(\ref{a1}).
However, the evolutions are completely different, which might be
regarded as dual in certain sense.

When ${\epsilon }\sim {1\over (\eta_*-\eta)^2}$ and $a$ is nearly
constant, one of the solutions is $H\simeq {1\over \alpha
t}+\Lambda_*$, in which $\alpha$ is constant and ${1\over \alpha
t}\ll \Lambda_*$ since initially $t\ll -1$. Thus $H$ is also
hardly changed for some times. This is adiabatic ekpyrosis given
in \cite{KS1}, see \cite{KM} for the detailed discussion for the
solutions and \cite{LMV} for criticism. Here though $k^2\ll
z^{\prime\prime}/z$, the perturbation mode is actually still
inside the Hubble horizon, since $k=aH$ and both $a$ and $H$ are
hardly changed. Thus a period after it is required to extend the
perturbation mode out of the Hubble horizon.


We, in this paper, will consider the evolution with ${|\epsilon |}
\sim (\eta_*-\eta)^4$ and constant $a$. In this case, the spectrum
of the curvature perturbation is induced by its increasing mode,
while in \cite{KM},\cite{JK}, the spectrum is induced by the
constant mode. This difference is significant. We will see that,
different from that in \cite{KM},\cite{JK}, here the perturbation
mode can naturally leave the horizon, and also there is not the
problem pointed out in \cite{LMV}.

We begin with ${|\epsilon |} \sim (t_*-t)^4$, since $a$ is nearly
constant which brings $\eta\sim t$. Thus by integral for it, we
have \be H\sim \pm\, {1\over \Lambda_*^4 (t_*-t)^5}, \label{H}\ee
where initially $t\ll -1$ and $|t|\gg |t_*|$, and $\Lambda_*\simeq
1/|t_*|$ is constant, which is regarded as the exit scale of $H$.
The positive solution, i.e. the expansion solution, is for
$\epsilon<0$. The minus, i.e. the contraction solution, is for
$\epsilon>0$. $a$ can be obtained by $\ln{a}= \int Hdt$, which is
\be \ln{a}\sim \pm\,{1\over \Lambda_*^4(t_*-t)^4}.\label{a}\ee
When initially $t\ll -1$, $a\simeq 1$ and $|H|$ is highly small
and negligible, see Eq.(\ref{H}), however, $|\epsilon | \simeq
\Lambda_*^4(t_*-t)^4\gg 1$ is not constant and will gradually
decrease with the evolution. When $t\simeq {\cal O}(1)t_*$, this
phase of background evolution ends. In this epoch, $a\simeq e$
implying $a$ is slowly expanding during this phase, or $1/e$
implying $a$ is slowly contracting, $|H|\sim |H_*|\simeq
\Lambda_*$ which is large, and $|\epsilon|\sim 1$. In principle,
it can be expected that after this slowly expanding or contracting
phase, the evolution of standard cosmology begins. This will bring
completely distinct scenarios of early universe. We list Tab.1,
which is a brief of above discussions.

The slowly evolving of $a$ and the rapidly increasing of $H$ lead
that the perturbation modes can be naturally extended out of
Hubble horizon during this phase. The efolding number for the
primordial perturbation generated during this phase is \be {\cal
N}\simeq \ln\left({H_*\over H}\right), \ee since $k=a|H|$ and $a$
is nearly constant. Thus in principle, the enough efolding number
can be obtained. The details of the model and the evolution of
perturbation mode are visualized in Fig.1, in which the initial
time $|t|\simeq 10|t_*|$ is set for simplicity.

\begingroup
\begin{table}
    \label{spectrum}
    \begin{tabular}{|c|c|c|c|c|}
      \hline
      \multicolumn{5}{|c|}{Nearly scale invariance of $\zeta$ (single field)}
       \\ \hline \hline   & \multicolumn{2}{|c|}{$|\epsilon|$ is slowly changed } &
       \multicolumn{2}{|c|}{$a$ is slowly changed } \\
      \hline \hline
      $C$  & \multicolumn{2}{|c|}{$a\sim t^n (n\gg 1)$} &
       \multicolumn{2}{|c|}{$|\epsilon|\sim {1\over (t_*-t)^2}$ (initially $|\epsilon|\simeq 0$)}
       \\
      \hline  \hline
      $D$  & \multicolumn{2}{|c|}{$a\sim (t_*-t)^{2\over 3}$ } &
       \multicolumn{2}{|c|}{$|\epsilon|\sim (t_*-t)^4$ (initially $|\epsilon|\gg 1$) }\\
      \hline \hline
    \end{tabular}
    \caption{The possibilities of nearly scale invariance of $\zeta$ for single field. The
    $C$ or $D$ denotes that $\zeta$ is dominated by its constant mode or increasing mode.
    When $|\epsilon|$ is slowly changed, $a\sim t^n (n\gg 1)$ is that of
    inflation \cite{G},\cite{LAS},\cite{S1},\cite{MC}, while
    $a\sim (t_*-t)^{2\over 3}$ is that of the contraction with $\epsilon\simeq 1.5$ or $w\simeq 0$ \cite{Wands99},\cite{FB},\cite{SS}.
    When $a$ is slowly changed, $|\epsilon|\sim {1\over
    (t_*-t)^2}$ implies that $|\epsilon|\sim 0$ initially and $|\epsilon|\gg
    1$ around the ending since $t<0$, which is that of adiabatic ekpyrosis given in \cite{KS1},\cite{KM}, while $|\epsilon|\sim (t_*-t)^4$ implies that
$|\epsilon|\gg 1$ initially and
    $|\epsilon|\simeq 1$ around the ending, which is that given in this paper. These listed here, for constant $c_s$, might be
    the simplest possibilities obtaining scale invariant $\zeta$. In principle,
    both $a$ and $|\epsilon|$ can be changed, however, the case is slightly
    intractable.}
  \end{table}
  \endgroup

\begin{figure}
\includegraphics[width=7cm]{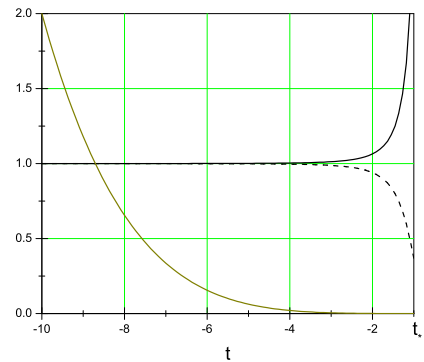}
\caption{ The black lines are the evolutions of $a$, and the solid
line and dashed line correspond to the slowly expanding and slowly
contracting, respectively. The dark yellow line is that of
$1/|H|$. The lines are plotted with Eqs.(\ref{H}) and (\ref{a}),
in which for simplicity $\Lambda=|t_*|=1$ is set and the initial
time is $t\simeq 10t_*$. We can see that the change of $a$ is not
negligible is only around $t\simeq 2t_*$, at this epoch the
exiting is assumed to occur. During this phase, due to the rapidly
change of $H$, the perturbation mode initially inside the horizon,
i.e.$\lambda\sim a\ll 1/|H|$, will naturally leave the horizon,
i.e.$\lambda\gg 1/|H|$.}
\end{figure}

When $k^2 \gg z^{\prime\prime}/z$, $u_k\rightarrow {1\over
\sqrt{2k}}e^{ik\eta}$, which gives the initial condition of mode
evolution. $z^{\prime\prime}/z$ is increased with the time. When
$k^2\ll z^{\prime\prime}/z$, the solution of Eq.(\ref{uk}) brings
\be {\cal P}^{1/2}_\zeta \simeq \sqrt{k^3}\left|{u_k\over
z}\right|\simeq {1\over |t_*|^3\Lambda_*^2}\simeq |H_*|,
\label{PP}\ee since $|H_*|\simeq 1/|t_*| \simeq \Lambda_*$ and $a$
is constant. In certain sense, ${\cal P}^{1/2}$ is determined by
$|H_*|$ around the exiting time is the reflection that on
superhorizon scale $\zeta$ is increased, since when it leaves the
horizon $|H|$ is quite small. The observations requires $|H_*|\sim
\Lambda_*\sim 10^{-5}$, since ${\cal P}^{1/2}\sim 10^{-5}$, which
implies that the scale around the exiting is about $10^{16}$Gev.
This is not any finetunning.

We will calculate the perturbation of energy density, following
\cite{LMV}. The equation of metric perturbation is same as
Eq.(\ref{uk}), however, here $u$ is defined as
$H\sqrt{|\epsilon|}u_k\simeq\Phi_k$ and $z$ is replaced with
$\theta=1/z$. Thus the solution is $u_k\sim \theta\int
{d\eta\over\theta^2 }$ or $\theta$ for $k^2\ll
\theta^{\prime\prime}/\theta$. That of increasing mode is $u\sim
\theta$, since $a$ is hardly changed and $z\sim
\sqrt{|\epsilon|}\sim \Lambda_*^2(t_*-t)^2$. Thus we have \be
\Phi_k\simeq {H\sqrt{|\epsilon|}\over z}\sim H. \ee The amplitude
of the energy density perturbation on large scale is given by $
({\delta\rho /\rho})_k \simeq \Phi_k+{{\dot \Phi}_k/ H}\sim {{\dot
\Phi}_k/ H}$, since here $\Phi_k\ll {{\dot \Phi}_k/ H}$ for
$\Lambda_*^4 (t_*-t)^4\ll 1$. Thus when the slow expanding phase
ends, $|H_*|\simeq 1/|t_*| \simeq \Lambda_*$, we approximately
have \be \left({\delta\rho \over \rho}\right)_k\sim {1\over
(t_*-t)} \simeq |H_*|,\ee  which is consistent with (\ref{PP}). In
\cite{LMV}, it has been pointed that the perturbations of the
energy density in \cite{KS1} can be too large invalidating the use
of the perturbation theory. The reason is that the amplitude of
the perturbation after leaving the horizon is constant, which is
too small to be responsible for the observations, thus an
astronomically large value of parameter $c$ is required to uplift
this amplitude \cite{KS1}. However, here since the amplitude of
the perturbation is increasing, at the end time of slow expansion
a suitable amplitude of perturbation can be naturally obtained.

The equation of motion of the tensor perturbation $h_k$ is that in
Eq.(\ref{uk}) with the replacements of $u_k\simeq ah_k$ and $z=a$,
in which $h_k$ is the $k$ mode of the tensor perturbation
$h_{ij}$, which satisfies $\delta^{ij} h_{ij}=0$ and $\partial^{i}
h_{ij}=0$, here $i$ and $j$ are the spatial index. When the
expansion or contraction is slow, the dominated term of
$a^{\prime\prime}/a$ is \be {a^{\prime\prime}\over a} \sim {1\over
\Lambda^4 (t_*-t)^6} \sim {{\cal O}(0)\over (\eta_*-\eta)^2} \ee
for $|t|\gg |t_*|$. Thus ${\cal P}_{T}^{1/2} \sim k$ is quite
blue, which implies that the amplitude of tensor perturbation is
negligible on large scale. This is the universal character of the
slowly evolving background, e.g.\cite{island}. Thus the detection
of tensor perturbation is significant for falsifying the slowly
expanding or contracting model.


We will briefly conceive how the required background evolution
might be obtained. We begin with the action with the negative
potential ${\Lambda^4}{\phi^3/ { M}^3}$, which will be expected to
bring the evolution of the slowly contracting, and the action with
negative kinetic energy and positive potential
${\Lambda^4}{\phi^3/ { M}^3}$, which will be expected to bring
that of the slowly expanding, in which $M$ and $\Lambda$ are the
constants of the mass dimensions. The slowly change of $a$
requires that initially $|\epsilon|\gg 1$, which implies $\rho\ll
|P|$ is negligible. $\rho\simeq 0$ brings ${\dot \phi}^2\simeq
\Lambda^4({\phi/ M})^3$. Thus ${\phi}\simeq {4M^3\over
\Lambda^4(t_*-t)^2}$
is obtained. Thus \be {\dot \phi}^2\simeq
\Lambda^4\left({\phi\over M}\right)^3 \sim {64 M^6\over
\Lambda^8(t_*-t)^6}, \label{X}\ee which is increased since
initially $t\ll -1$. We have \be {\dot H}\sim \pm\,{M^6\over
\Lambda^8(t_*-t)^6}, \label{hdot}\ee  since ${\dot H}\simeq -P$,
which by the integral will induce Eq.(\ref{H}), in which
$\Lambda_*^4\sim M^6/ \Lambda^8$. This result implies that, for
the slowly contracting or expanding, we have to add a term like
$(\partial\phi)^{2/3}\Box\phi$ in the action, which will assure
\be\rho\sim H^2\sim {1\over (t_*-t)^{10}}\ee is negligibly small
but increased, since in this case $\rho\simeq H{\dot
\phi}^{5/3}\sim {1\over (t_*-t)^{10}}$. In certain sense, $\phi$
might be a Galileon \cite{NRT}, see its nontrivial generalization,
e.g.kinetic braiding \cite{DPSV} or \cite{KYY}. $|\epsilon |\simeq
{M^6(t_*-t)^4\over\Lambda^8}\gg 1 $ is gradually decrease. When
$t\simeq {\cal O}(1)t_*$, $|\epsilon |\simeq 1$, and $\rho$ has
become not negligible. Thus (\ref{hdot}) is not any more right
around this epoch, which signals the end of the slowly contracting
or expanding phase. Thus in principle, we can design the required
background evolution.


In certain sense, that of slow contraction might be only a simple
change of the adiabatic ekpyrotic scenario, in which a different
time dependence of $\epsilon$ is selected. Its jointing with the
standard cosmology requires a bounce mechanism, like in ekpyrotic
scenario \cite{KOS}, or \cite{GV},\cite{V}, or quintom bounce
\cite{Cai0704}. However, that of slow expansion is completely
different, in which the bounce is not required, and the jointing
of the slowly expanding with standard cosmology is only simply
reheating, like in phantom inflation \cite{PZhang}, since the
universe expands all along. However, since $\epsilon <0$, there is
ghost instability for $\zeta$. However, it can be thought that the
evolution with $\epsilon<0$ might be only the approximative
simulation of a fundamental theory below certain physical cutoff
during certain period, which is generally not Lorentz invariant
\cite{CJM}, and the full action should be ghost free. We will
provide the details of the model building of slowly expanding
scenario in the coming work \cite{Piao1105}, in which the
evolution of background satisfies the required conditions and
$c_s^2$ is nearly constant, and there is not the ghost
instability.

In conclusion, we have brought a possibility generating the scale
invariant spectrum, by which a viable scenario of early universe
might be implemented. In general, for single field, the scale
invariant spectrum of curvature perturbation can be given by
either its constant mode or its increasing mode. When $|\epsilon
|$ is rapidly changed while $a$ is slowly expanding or
contracting, the scale invariant spectrum of curvature
perturbation can be induced by its increasing mode. The
perturbation mode during this slowly evolving can be naturally
extended out of horizon, which is distinguished with that of
adiabatic ekpyrosis \cite{KS1}, in which the spectrum of curvature
perturbation is given by its constant mode. Here $c_s$ is constant
is set for simplicity, actually its change will enlarge the space
of solutions of the scale invariance of curvature perturbation
\cite{Picon},\cite{Piao0609},\cite{JM},\cite{KP},\cite{Kinney}.

\textbf{Acknowledgments} This work is supported in part by NSFC
under Grant No:10775180, 11075205, in part by the Scientific
Research Fund of GUCAS(NO:055101BM03), in part by National Basic
Research Program of China, No:2010CB832804.


\begin{thebibliography}{99}

\bibitem{Muk} V.F. Mukhanov, JETP lett. 41, 493 (1985); Sov. Phys. JETP. 68,
1297 (1988).

\bibitem{KS} H. Kodama, M. Sasaki, Prog. Theor. Phys.
Suppl. 78 1 (1984).

\bibitem{Mukhanov} V. Mukhanov, ``Physical Foundations of
Cosmology", (Cambridge University Press, 2005).


\bibitem{G} A. Guth, Phys. Rev. \textbf{D23},347(1981).

\bibitem{LAS} A.D. Linde, Phys. Lett. \textbf{B108}, 389 (1982); A.J. Albrecht and P. J.
Steinhardt, Phys. Rev. Lett. \textbf{48}, 1220 (1982).

\bibitem{S1} A.A. Starobinsky, Phys. Lett. \textbf{B91}, 99 (1980).


\bibitem{MC} V. Mukhanov and G. Chibisov, JETP \textbf{33} 549
(1981); A. Guth and S.Y. Pi, Phys. Rev. Lett. \textbf{49}, 1110
(1982); S.W. Hawking, Phys. Lett. \textbf{B115},295 (1982); A.A.
Starobinsky, Phys. Lett. \textbf{B117} 175 (1982); J.M. Bardeen,
P.J. Steinhardt and M.S. Turner, Phys. Rev. \textbf{D28} 679
(1983).

\bibitem{Wands99} D. Wands, Phys. Rev. \textbf{D60}, 023507
(1999).

\bibitem{FB} F. Finelli, R. Brandenberger, Phys. Rev. \textbf{D65}, 103522
(2002).

\bibitem{SS} A.A. Starobinsky, JETP Lett. \textbf{30}, 682 (1979).


\bibitem{KOS} J. Khoury, B. A. Ovrut, P. J. Steinhardt and N. Turok, Phys.
Rev. \textbf{D64}, 123522 (2001); Phys. Rev. \textbf{D66}, 046005
(2002).

\bibitem{PZhou} Y.S. Piao and E Zhou, Phys. Rev. \textbf{D68},
083515 (2003).

\bibitem{island} Y.S. Piao, Phys. Rev. \textbf{D72}, 103513
(2005); Phys. Lett. \textbf{B659}, 839 (2008); Phys. Rev.
\textbf{D79}, 083512 (2009).

\bibitem{LST} L.A. Boyle, P.J. Steinhardt and N. Turok, Phys. Rev. \textbf{D70},
023504 (2004).

\bibitem{Piao0404} Y.S. Piao, Phys. Lett. \textbf{B606}, 245 (2005); Y.S. Piao, Y.Z. Zhang, Phys. Rev. \textbf{D70}, 043516
(2004).

\bibitem{Lid} J.E. Lidsey, Phys. Rev. \textbf{D70}, 041302 (2004).



\bibitem{DH} D.H. Lyth, Phys. Lett. \textbf{B524}, 1 (2002);
Phys. Lett. \textbf{B526}, 173 (2002); R. Durrer and F. Vernizzi,
Phys. Rev. \textbf{D66}, 083503 (2002); S. Tsujikawa, R.
Brandenberger, F. Finelli, Phys. Rev. \textbf{D66}, 083513 (2002).

\bibitem{KS1} J. Khoury, P.J. Steinhardt, Phys. Rev. Lett. \textbf{104}, 091301
(2010).

\bibitem{KM} J. Khoury, G.E.J. Miller, arXiv:1012.0846.

\bibitem{JK} A. Joyce, J. Khoury, arXiv:1104.4347.

\bibitem{LMV} A. Linde, V. Mukhanov, A. Vikman, JCAP \textbf{1002}, 006
(2010).

\bibitem{NRT} A. Nicolis, R. Rattazzi, E. Trincherini, Phys. Rev. \textbf{D79},
064036 (2009).

\bibitem{DPSV} C. Deffayet, O. Pujolas, I. Sawicki, A. Vikman, JCAP \textbf{1010}, 026
(2010).

\bibitem{KYY} T. Kobayashi, M. Yamaguchi, J. Yokoyama, Phys. Rev. Lett. \textbf{105}, 231302
(2010).

\bibitem{GV} M. Gasperini and G. Veneziano, Astropart. Phys.
\textbf{1} 317 (1993).

\bibitem{V} M. Gasperini, G. Veneziano, Phys. Rept. \textbf{373}, 1 (2003); J.E. Lidsey, D. Wands
and E.J. Copeland, Phys. Rept. \textbf{337} 343 (2000).


\bibitem{Cai0704} Y.F. Cai, T. Qiu, Y.S. Piao, M.Z. Li, X.M.
Zhang, JHEP \textbf{0710}, 071 (2007).

\bibitem{PZhang} Y.S. Piao, Y.Z. Zhang, Phys. Rev. \textbf{D70}, 063513 (2004).



\bibitem{Piao1105} Z.G. Liu, J. Zhang, Y.S. Piao, arXiv:1105.5713.


\bibitem{CJM} J.M. Cline, S. Jeon, G.D. Moore, Phys. Rev. \textbf{D70} 043543
(2004).

\bibitem{Picon} C. Armendariz-Picon and E. A. Lim, JCAP \textbf{0312}, 002 (2003);
C. Armendariz-Picon, JCAP \textbf{0610}, 010 (2006)

\bibitem{Piao0609} Y. S. Piao, Phys. Rev. \textbf{D75}, 063517
(2007); Phys. Rev. \textbf{D79}, 067301 (2009)

\bibitem{JM} J. Magueijo, Phys. Rev. Lett. \textbf{100}, 231302 (2008).

\bibitem{KP} J. Khoury, F. Piazza, JCAP \textbf{0907}, 026 (2009).

\bibitem{Kinney} D. Bessada, W.H. Kinney, D. Stojkovic, J. Wang, Phys. Rev. \textbf{D81}, 043510
(2010).

\end{thebibliography}
\end{document}